\documentclass[pra,twocolumn,preprintnumbers,amsmath,amssymb,superscriptaddress,showpacs]{revtex4-1}

\usepackage{bm}
\usepackage{amsmath}
\usepackage{amsfonts}
\usepackage{amsthm}
\usepackage{graphicx}
\usepackage[caption=false]{subfig}
\usepackage[dvipsnames]{xcolor}
\usepackage{comment}
\usepackage[pdfstartview=FitH]{hyperref}
\usepackage[figure]{hypcap}
\usepackage{bbm}
\usepackage{enumitem}
\usepackage[normalem]{ulem}
\definecolor{myurlcolor}{rgb}{0,0,0.7}
\definecolor{myrefcolor}{rgb}{0.8,0,0}
\usepackage{hyperref}
\hypersetup{colorlinks, linkcolor=myrefcolor,
citecolor=myurlcolor, urlcolor=myurlcolor}
\usepackage[draft]{fixme}

\newcommand\ot{\otimes}
\newcommand{\<}{\langle}
\renewcommand{\>}{\rangle}
\newcommand{\Tr}{\text{Tr}}

\def\beq{\begin{equation}}
\def\eeq{\end{equation}}
\def\be{\begin{equation}}
\def\ee{\end{equation}}
\def\ben{\begin{eqnarray}}
\def\een{\end{eqnarray}}
\def\beqa{\begin{eqnarray}}
\def\eeqa{\end{eqnarray}}
\def\eea{\end{array}}
\def\bea{\begin{array}}
\newcommand{\bei}{\begin{itemize}}
\newcommand{\eei}{\end{itemize}}
\newcommand{\bee}{\begin{enumerate}}
\newcommand{\eee}{\end{enumerate}}

\begin{document}

\title{Closed timelike curves and the second law of thermodynamics}

\author{Ma\l{}gorzata Bartkiewicz}
\affiliation{Faculty of Physics, Adam Mickiewicz University, 61-614 Pozna\'{n}, Poland}

\author{Andrzej Grudka}
\affiliation{Faculty of Physics, Adam Mickiewicz University, 61-614 Pozna\'{n}, Poland}

\author{Ryszard Horodecki}
\affiliation{Institute of Theoretical Physics and Astrophysics, Faculty of Mathematics,
Physics and Informatics, University of Gdansk, 80-308 Gda\'{n}sk, Poland, National Quantum Information Centre of  Gda\'{n}sk, 81-824 Sopot, Poland}

\author{Justyna \L{}odyga}
\affiliation{Faculty of Physics, Adam Mickiewicz University, 61-614 Pozna\'{n}, Poland}

\author{Jacek Wychowaniec}
\affiliation{School of Materials, The University of Manchester, Oxford Road, M13 9PL, Manchester, UK}
\affiliation{Manchester Institute of Biotechnology, The University of Manchester, 131 Princess Street, M1 7DN, Manchester, UK}

\date{\today}
\begin{abstract}

One out of many emerging implications from solutions of Einstein's general relativity equations are closed timelike curves (CTCs), which are trajectories through spacetime that allow for time travel to the past without exceeding the speed of light. Two main quantum models of computation with the use of CTCs were introduced by Deutsch (D-CTC) and by Bennett and Schumacher (P-CTC). Unlike the classical theory in which CTCs lead to logical paradoxes, the quantum D-CTC model provides a solution that is logically consistent due to the self-consistency condition imposed on the evolving system, whereas the quantum P-CTC model chooses such solution through post-selection. Both models are non-equivalent and imply nonstandard phenomena in the field of quantum computation and quantum mechanics. In this work we study the implications of these two models on the second law of thermodynamics -- the fundamental principle which states that in an isolated system the entropy never decreases. In particular, we construct CTC-based quantum circuits which lead to decrease of entropy.

\end{abstract}


\maketitle

\section{Introduction}
A concept of voyage through time has been puzzling modern physicists for a long time. 
Einstein's general theory of relativity allows the existence of closed timelike curves (CTCs) \cite{Godel1949,Thorne1988}, where an object could travel back in time and interact with a former version of itself.

The possible existence of CTCs points to a variety of logical paradoxes, such as famous grandfather paradox \cite{Nahin1999}. However, such paradoxes can be efficiently eliminated in quantum theory. One of the models that does so was proposed by Deutsch \cite{Deutsch1991}. His model of CTC (D-CTC) operates within the density matrix formalism to describe the states of the two registers involved: a chronology-respecting (CR) system and a CTC chronology-violating (CV) quantum system, which interact with each other \textit{via} some unitary operation. For any such unitary operation and state of the CR system, a self-consistency condition \cite{Novikov1990} must be satisfied, where the state of the CV system prior to and after the interaction is set to remain the same. Such condition ensures the exclusion of the arising grandfather-like paradoxes, but also introduces a nonlinear evolution. This, in turn, gives rise to peculiar phenomena, e.g., violation of no-cloning theorem \cite{Ahn2010,BrunWinter2013}, increasing entanglement with LOCC \cite{Moulick2016}, and distinguishing non-orthogonal quantum states \cite{BrunWilde2009} which has been experimentally simulated in \cite{Ralph2014}. In addition, there exist another unusual implications of this model concerning the possible enhancement of power of D-CTC-assisted computation, such as the equivalence of classical and quantum computing or the ability to efficiently solve NP-complete and PSPACE-complete problems \cite{Brun2003,Bacon2004,Aaronson2009}.

Bennett \textit{et al.} \cite{Bennett2009} questioned all of above implications showing that they simply stem from ``falling into a linearity trap", i.e., generalization of the analysis made for pure input states to their mixture. Due to the classical ignorance one can have about the preparation of a state, the authors considered as an input state a classically-labeled unknown mixture of states, where the labels are encoded in orthogonal states of some reference system. When analyzing the evolution of such mixture in the Deutsch's model, one gets that the output of a D-CTC-assisted circuit may become independent of the input data, in particular, all the correlations between CR system and the reference system can be lost. This comes from the nonlinear evolution, where the map acting on the mixture of states is not equivalent to a mixture of states individually acted upon this map. Based on this argument, Bennett \textit{et al.} concluded that the use of D-CTC does not entail any computational benefits and that observation of the phenomena against quantum postulates is blurred. However, the work of Bennett \textit{et al.} \cite{Bennett2009} was undermined by other authors \cite{RalphMyers2010,Cavalcanti2010}, and the correct approach to the computation in D-CTC model still remains debatable.

Different idea for a computational model of time travel which uses post-selected teleportation (P-CTC) came forth initially from Bennett and Schumacher \citep{Bennett2005}, and was subsequently developed in \citep{Lloyd2011PRL,Lloyd2011PRD,Svetlichny2011}. In a P-CTC model, the evolution of a time-traveling quantum state is described by a well-known, but modified, quantum teleportation protocol without communication. Namely, standard Bell measurement is replaced by a projection into a fixed state $|\Phi^+\rangle$ and renormalization of probabilities. The renormalization proceeds as follows: if the projection does yield identically zero, then the output is not defined and the evolution is lost; if the probability of a desired outcome is nonzero, it is renormalized to 1. The first rule allows to avoid logical paradoxes, and occurs for a set of input states of measure 0. Note that in D-CTC model the consistent solutions exist for any given input CR states.

One of the causes for the arising peculiar phenomena against quantum mechanics in Deutsch's model was a nonlinear evolution. Likewise, in a CTC model with post-selection we deal with a nonlinear process, however of a different nature than in D-CTC, i.e., nonlinearity comes from the renormalization of the probabilities. In this manner, Brun and Wilde  \cite{BrunWilde2011} reformulated the ``linearity trap" argument of  Bennett \textit{et al.} for a P-CTC model. The authors considered three distinct quantum state representations due to their various preparation procedures. First, so called proper mixture, corresponds to an ensemble of pure states. Second, improper mixture, describes a subsystem of an entangled system. And last, fundamental mixture, that covers all other interpretations. As authors state, these mixtures remain indistinguishable in a linear quantum mechanics. In a nonlinear regime, however, one obtains different results depending on the origin of an input state (i.e., the type of a mixture). On the other hand, according to \cite{BrunWilde2011,RalphMyers2010,Cavalcanti2010}, it seems that Bennett's \textit{et al.} argument \cite{Bennett2009}, which is based on choosing a labeled mixture as a proper input state, does not differentiate between the different possible interpretations of a state.

Brun and Wilde \cite{BrunWilde2011} also examined some of the key implications of P-CTC model, namely, the ability to distinguish linearly independent quantum states, as well as the potential use of a single P-CTC qubit to complete different computational tasks such as effective integers factorization or solving NP-problems. It has also been shown that the experimental realization of this model is possible \cite{Lloyd2011PRL} due to a post-selection procedure involved.

The debate regarding which model, either D-CTC or P-CTC, if any, is correct is still ongoing \cite{Ralph2007, Ralph2009,Wallman2012,Cavalcanti2010,RalphMyers2010,Allen2014,DongChenZhou2017}. In this work we consider aforementioned models of CTCs: D-CTC and P-CTC, and explore their behavior with respect to one of the most elementary laws of physics: the second law of thermodynamics. In particular, we study the entropy change of the CR quantum system that undergoes transformation in different CTC-assisted quantum circuits. We start by discussing the D-CTC model in detail and then show that it allows the violation of the second law of thermodynamics, when one ``falls into the linearity trap" of Bennett \textit{et al.} As next proved, this is no longer a case when the linearity trap is avoided, namely when the evolution is performed on the initial mixture of states rather then its components. In this case, the entropy of evolving CR quantum system decreases. Then, we recall the Bennett's and Schumacher's model of post-selected CTCs and show that for almost any mixture it implies the violation of the second law of thermodynamics. We summarize by interpreting our results in the context of a long lasting debate between the enthusiasts of D-CTC and P-CTC model.

\begin{figure}
\centering
\includegraphics[width=.45\textwidth]{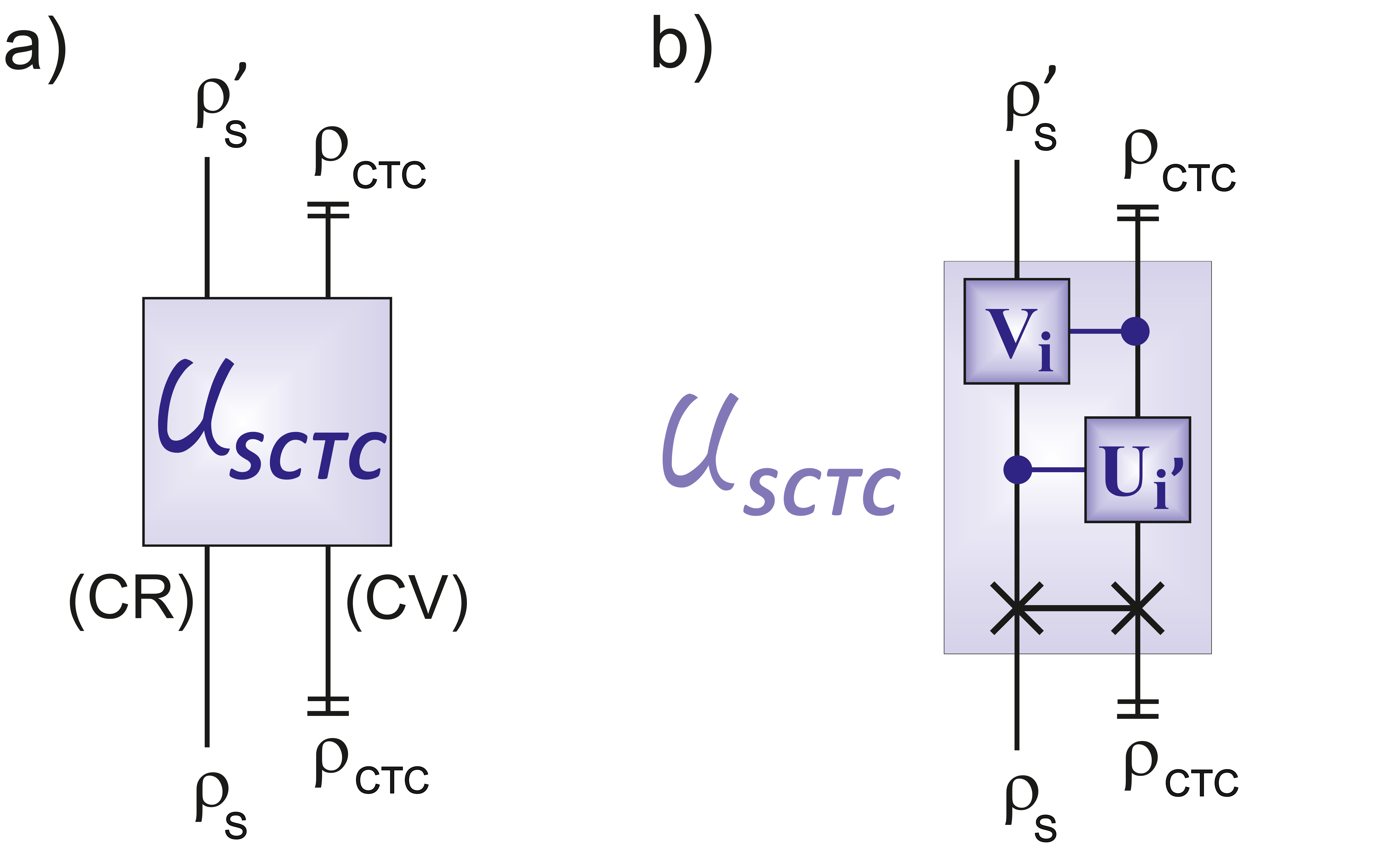}
\caption{ a) D-CTC quantum circuit, where two input states: $\rho_S$ and $\rho_{CTC}$ interact with each other \textit{via} a unitary operation $U_{SCTC}$. The self-consistency condition enforces the output state $\rho_{CTC}$ of a qudit traveling along CTC to match the corresponding input state $\rho_{CTC}$. The output state $\rho'_{S}$ of a CR qudit continues its travel into an unambiguous future; b) D-CTC circuit that allows to violate the second law of thermodynamics. The unitary operation $U_{SCTC}$ is chosen as a product of SWAP, controlled-$U_{i'}$ and controlled-$V_{i}$ operations.}
\label{ctcdeutsch}
\end{figure}

\section{D-CTC AND THE SECOND LAW OF THERMODYNAMICS}
\subsection{Preliminaries to a D-CTC model}
The schematic of a CTC-assisted quantum circuit based on Deutsch's model is presented in Fig. \ref{ctcdeutsch} a). It covers two spacetime regions: the one that preserves the time chronology (CR), and the other where the chronology is violated (CV). The input $\rho_S$ (the density matrix of a quantum system S) that enters the former region from the unambiguous past, interacts with the input $\rho_{CTC}$ (density matrix of a CTC system) existing in the latter region, and afterwards, it continues its travel into unambiguous future.

Before interaction, the systems are treated as separated and therefore the initial state is described by product of density matrices:
\be\label{productstate}
\rho_{SCTC}=\rho_S \ot \rho_{CTC}.
\ee
The interaction set by a unitary operation $U_{SCTC}$ causes the entering system to evolve into a state
\be
\label{evolution}
\rho'_{SCTC}=U_{SCTC} (\rho_S \ot \rho_{CTC}) U_{SCTC}^{\dagger}.
\ee

Now, one imposes the self-consistency condition \citep{Deutsch1991} 
\be
\label{selfcons}
\rho_{CTC}=\Tr_{S} (\rho'_{SCTC})
\ee
on the final state $\rho_{CTC}$ of a CTC qudit, that results from tracing out the system $\rho'_{SCTC}$ defined in Eq. \eqref{evolution} over subsystem S. Note that the self-consistency condition \eqref{selfcons} enforces the initial and final state of the CTC qudit, that appear respectively on the right-hand side (RHS) of Eq. \eqref{evolution} and left-hand side (LHS) of Eq. \eqref{selfcons}, to be the same. Deutsch showed that the fixed point of transformation \eqref{selfcons} always exists.

After interaction, the state of the CR qudit
\be
\label{finalSstate}
\rho'_{S}=\Tr_{CTC} (U_{SCTC} (\rho_S \ot \rho_{CTC}) U_{SCTC}^{\dagger})
\ee
is a nonlinear function of the initial state of system $\rho_S$, since it depends on both: $\rho_S$ itself and $\rho_{CTC}$ \eqref{selfcons} that also depends on $\rho_S$.

\subsection{Can the second law be violated with D-CTC?}

Now, we will use above model of D-CTC to examine the entropy change of a state of a qudit traveling along the CR trajectory. It remains an intriguing and difficult question in itself of what physical outcome one should expect. Indeed, in a CTC-assisted quantum model the time evolution is rather peculiar: in one region the system moves forward in time, whilst another system moves forward and backward in time. Additionally, quantum systems moving within both regions are allowed to interact and make an impact on each other.

According to Deutsch \cite{Deutsch1991}, the chronology violation does not affect the validity of the second law of thermodynamics, which still applies. However, as we show in the next paragraph, with an appropriate choice of an unitary operation $U_{SCTC}$, one can cause  the entropy of a chronology-respecting state $\rho_S$ to decrease. In our calculations of the entropy change, the starting and end points are taken when the only object in the system is qudit S. This is possible since CTC exists only for a finite period of time, and therefore it does not contribute to the entropy of a system in these chosen moments. Also, we assume that the state of the surrounding environment does not change for these points.

When working within a D-CTC model framework, one has to be aware of Bennett's \textit{et al.} ``linearity trap" argument. Therefore, for a complete analysis, we begin with a mixed state $\rho_S$ and study the problem using two approaches. Firstly, in the spirit of \cite{BrunWilde2009}, we evolve each component of a mixture separately and then average over obtained evolutions. Speaking a language of Bennett \textit{et al.}, we let ourselves ``fall into a linearity trap". In the second approach, the mixed state is evolved as a whole, and the ``linearity trap" is avoided. Interestingly, these two approaches lead to very different conclusions.

\subsection{Approach 1: Violation of the second law of thermodynamics in a D-CTC model}\label{DCTCapproach1}
Let us consider a proper mixture, i.e,  a statistical ensemble $\{p_k,|\psi_k\>\}$ of states $|\psi_k\rangle$ with respective probabilities $p_k$, which gives the initial state of a $d$-dimensional system S in the form:
\be\label{mixedstate}
\rho_S=\sum_k p_k |\psi_k\rangle \langle \psi_k|.
\ee
In our first approach we choose the interaction between S and CTC systems in such a way to intentionally violate the second law of thermodynamics. In particular, we show how to construct a D-CTC-assisted quantum circuit that transforms the initial state \eqref{mixedstate} into a pure state $|\varphi\>$. The above transformation will be examined by evolving each component $\{|\psi_k\>\}$ of a mixture \eqref{mixedstate} separately and then averaging over obtained evolutions.

To this end, let us first choose the unitary operation as
\be
\label{unitaryDeutsch}
U_{SCTC}\!\!=\!\!\left(\!\sum_{i} V_{i} \otimes |i\rangle \langle i| \right) \!\! \left(\!\sum_{i'}|i'\rangle \langle i'| \otimes U_{i'} \!\right) \text{SWAP},
\ee
where the first two components are controlled-$V_{i}$ and controlled-$U_{i'}$ operations, respectively, and $\text{SWAP}= \sum_{j,j'}|j\rangle \langle j'| \otimes |j'\rangle \langle j|$. An associated CTC quantum circuit is shown in Fig. \ref{ctcdeutsch} b). 

Next, we need to define both sets of unitary transformations $\{U_k \}$, $\{V_k\}$ from Eq. \eqref{unitaryDeutsch} such that the initial mixed state is mapped to a pure state and that the fixed point of the self-consistency condition \eqref{selfcons} is unique. Therefore, we choose $U_k$ and $V_k$ to satisfy the following equations:
\ben
\label{opU}
U_{k} |\psi_{k}\> &=& |k\>,\\
\label{opV}
V_{k} |k\> &=&  |\varphi\>,
\een
for each $k$. In Appendix \ref{APPENDIXselfcons} we prove that such choice guarantees the existence of the unique self-consistent solution $\rho_{CTC}=|k\rangle \langle k|$. Now, our approach is to study an evolution of the component of an initial proper mixture  instead of a whole state \eqref{mixedstate}.  And so, the componentwise evolution of the initial state of the whole system throughout the circuit reads as:
\ben
\label{transformation_calculations}
\nonumber
&& U_{SCTC} (|\psi_{k}\> \ot  |k\> ) \\
\nonumber
&=&\!\!\left(\!\sum_{i} V_{i} \!\otimes\! |i\rangle \langle i| \!\right)\!\!  \left(\!\sum_{i'}|i'\rangle \langle i'| \!\otimes\! U_{i'}\! \right)\! (\text{SWAP}) (|\psi_{k}\>\! \ot \! |k\> )\\
\nonumber
&=&\!\!\left(\!\sum_{i} V_{i} \!\otimes\! |i\rangle \langle i| \!\right)\!\!  \left(\!\sum_{i'}|i'\rangle \langle i'| \!\otimes\! U_{i'}\! \right) ( |k\> \ot |\psi_{k}\>)\\
\nonumber
&=& \left(\!\sum_{i} V_{i} \!\otimes\! |i\rangle \langle i| \!\right) (|k\> \ot  |k\> )\\
&=& |\varphi\> \ot  |k\>,
\een
where in the third equality we used Eq. \eqref{opU}, and in the last equality Eq. \eqref{opV}.

Next step is to average over obtained evolutions, i.e., $\sum_k p_k (U_{SCTC} (|\psi_{k}\>\<\psi_{k}| \ot  |k\> \<k |))=\sum_k p_k  (|\varphi\>\<\varphi| \ot  |k\> \<k |,$ where we used Eq. \eqref{transformation_calculations}. One can note that by tracing out the obtained state over subsystem CTC, one obtains the pure state $|\varphi\>$ as an output state of the system S, as desired.

Let us now examine the change of entropy of the state of CR system due to the above transformation. Before interaction, the von Neumann entropy $S(\sum_{k} p_k|\psi_k\> \<\psi_k|)$ of the initial mixed state of the qudit S is nonzero. Since the entropy of a final pure state is zero, $S\left(|\varphi\> \<\varphi| \right) =0$, we observe the decrease of entropy which indicates the violation of the second law of thermodynamics.

To sum up, in this approach we intentionally ``fell into a linearity trap" of Bennett's \textit{et al.} by making an analysis only for pure states that compose an initial proper mixture, and then inferring about their mixture. This way, with an appropriate choice of quantum circuit operations we were able to violate the second law. Clearly, now a question arises: how would the entropy of the CR state change when the initial mixed state evolves from the beginning as a whole. In the next paragraph we present an approach to this issue. In particular, we purposely avoid ``falling into a linearity trap" and show that in such case the violation of the second law is not possible regardless of the nature of an input state as well as the unitaries used.

\subsection{Approach 2: Conservation of  the second law of thermodynamics in a D-CTC model}\label{DCTCapproach2}
In this approach we avoid so called ``linearity trap" introduced by Bennett \textit{et al.} As an input CR state we choose either improper or fundamental mixed state $\rho_S$, and examine how it evolves in the D-CTC quantum circuit under a unitary interaction. As it turns out, in such case the entropy never decreases. Below a reformulation of the Deutsch's proof \cite{Deutsch1991} which uses the notion of mutual information is presented \footnote{We thank an anonymous referee for providing us this proof.}.

Before interaction, the states of qudits S and CTC are assumed to be completely uncorrelated \eqref{productstate}, and therefore the quantum mutual information $I$ of an initial system SCTC is zero:
\be\label{mutual1}
I(\rho_{SCTC}) \equiv S(\rho_S)+S(\rho_{CTC})-S(\rho_{SCTC}) = 0,
\ee
where $S$ denotes the von Neumann entropy.
After interaction, the quantum mutual information is given by:
\be\label{mutual2}
I(\rho'_{SCTC}) \equiv S(\rho_S')+S(\rho_{CTC})-S(\rho'_{SCTC}),
\ee
with the self-consistency condition \eqref{selfcons}: $\rho'_{CTC}=\rho_{CTC}$ being fulfilled. Now, since the mutual information is always non-negative:
\be\label{mutualnonnegativity}
I(\rho'_{SCTC}) \geq 0
\ee
one obtains:
\be
\label{mutualproof1}
S(\rho_S')-S(\rho'_{SCTC}) \geq S(\rho_S)-S(\rho_{SCTC}),
\ee
which combines Eq. \eqref{mutualnonnegativity} with Eqs. \eqref{mutual1} and \eqref{mutual2}. Using the fact that interaction is unitary (and thus does not change the entropy $S(\rho'_{SCTC})=S(\rho_{SCTC}))$,  Eq. \eqref{mutualproof1} simplifies to:
\be
\label{mutualproof2}
S(\rho_S') \geq S(\rho_S),
\ee
 and so the entropy of the CR system does not decrease. Indeed, for improper and fundamental mixtures the above approach yields the second law of thermodynamics always holds. However, there is no clear agreement on how the proper mixtures interact with CTCs. There can be two cases: 1) either they interact with CTCs as a whole and therefore they follow exactly the same approach as the improper and fundamental mixtures described above, or 2) the evolution runs for individual components of a mixture (which corresponds to the approach 1 from Sec. \ref{DCTCapproach1}). It remains questionable whether CTCs discriminate between proper and improper mixtures and therefore it cannot be definitely concluded whether D-CTC would lead to violation or conservation of the second law of thermodynamics. 

\begin{figure}
\centering
\includegraphics[width=.45\textwidth]{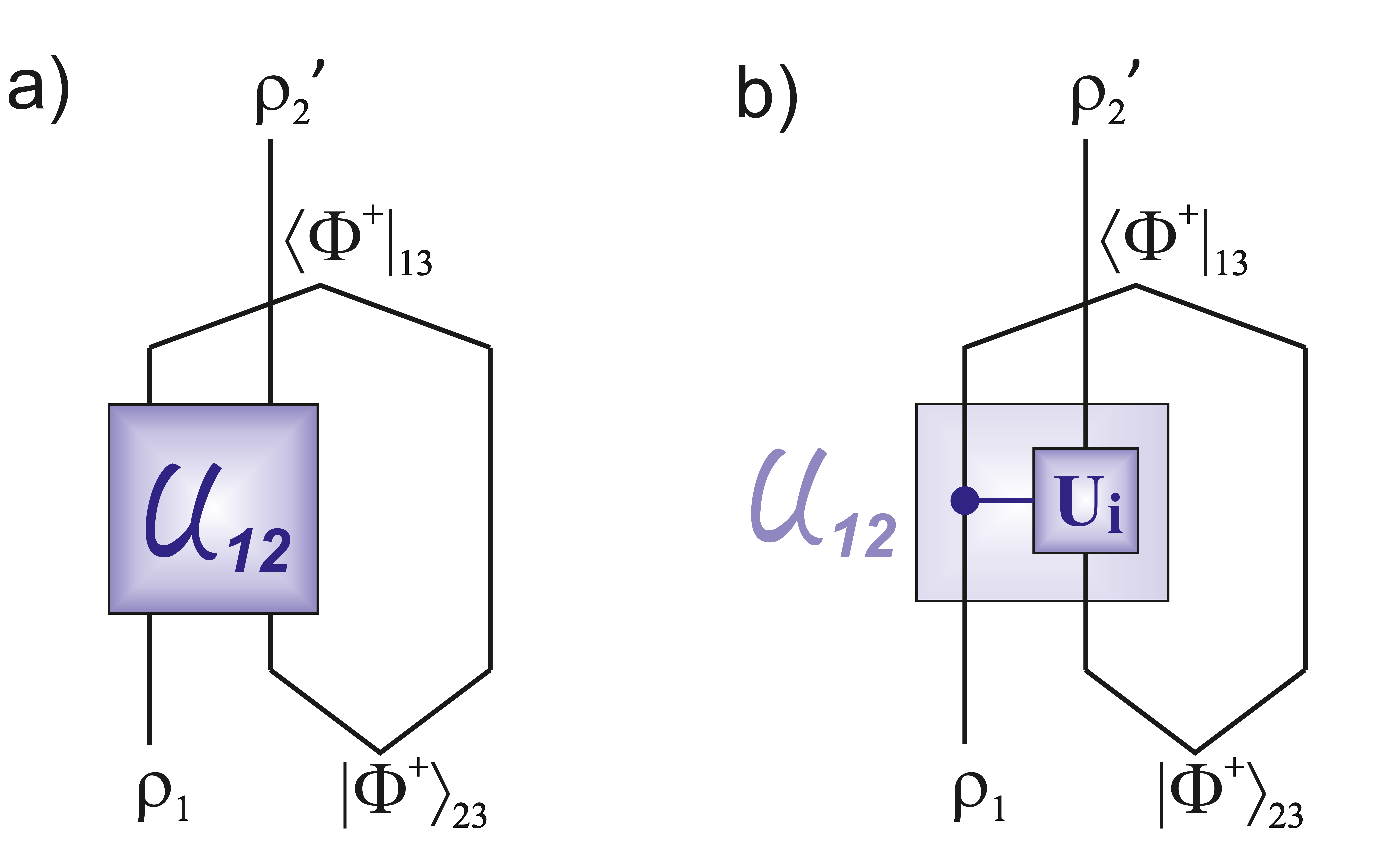}
\caption{ a) P-CTC quantum circuit, where the unitary operation $U_{12}$ acts on a qudit 1 in a state $\rho_1$ and qudit 2 from a maximally entangled pair in the state $|\Phi^+\>_{23}$. Afterwards, qudits 1 and 3 are projected onto the state $\<\Phi^+|_{13}$ with probability 1, and a state $\rho'_2$  of qudit 2 is post-selected; b) P-CTC circuit that allows to violate the second law of thermodynamics. The unitary interaction is chosen in a form of controlled--$U_{i'}$ operation.}
\label{ctcrho1i2}
\end{figure}

\section{P-CTC and the second law of thermodynamics}
\subsection{Preliminaries to a P-CTC model}

The post-selected closed timelike curve (P-CTC) \cite{Bennett2005} is presented in Fig. \ref{ctcrho1i2} a). In the P-CTC-assisted circuit there are 3 inputs: qudit 1 in a state $\rho_1$, and a pair of qudits (2 and 3) in a maximally entangled state $|\Phi^+\rangle_{23}$. This results in an overall initial state $\rho$:
\be\label{statepctc}
\rho=\rho_1 \ot |\Phi^+\rangle_{23}\langle \Phi^+|_{23}.
\ee
Qudit $1$ interacts with a qudit $2$ from the entangled pair through the unitary operation $U_{12}$. After interaction,  a state of qudits 1 and 3 is projected onto a maximally entangled state $|\Phi^+\rangle_{13}$ with probability 1, unlike an ordinary quantum measurement. The output of the circuit is a state $\rho'_2$ of qudit 2. Before renormalization it reads as:
\be\label{projectionpctc}
\tilde{\rho}'_2 = \< \Phi^+|_{13} U\rho  U^{\dagger} |\Phi^+\>_{13},
\ee
where  $U=U_{12} \ot I_3$ and $\rho$ is given in Eq. \eqref{statepctc}. 

The above state \eqref{projectionpctc} is a result of a nonlinear evolution.  This is due to a post-selection where the outcome is chosen with certainty (because of renormalization). Therefore, in a P-CTC model the issue of nonlinearity remains, however of a different nature to that in a D-CTC. Brun and Wilde reconsidered a  ``linearity trap" problem in P-CTC \cite{BrunWilde2011} and stated that it is inherently associated with a different characterizations of evolving states: proper, improper or fundamental mixtures, which are no longer equivalent in a nonlinear regime. When Bennett \textit{et al.} presented the ``linearity trap" argument \cite{Bennett2009}, they seemed to treat labeled mixtures as fundamental mixtures, which was later criticized by other authors \cite{Cavalcanti2010,RalphMyers2010}. The same problem arises in the presence of a post-selected CTC. Depending on the approach, one can obtain utterly different results. When initial state is described as a proper mixture and the components of that mixture are evolved separately and subsequently renormalized, the original weights in the mixture remain unaffected. However,  for the evolution of a whole mixed state (i.e., when the initial state is treated as a fundamental mixture), the original weights may become updated and dependent on the full mixture. Let us recall that the first case corresponds to our first approach to D-CTC and the second law of thermodynamics in Sec. \ref{DCTCapproach1}, and the other to the second approach in Sec.  \ref{DCTCapproach2}.

\subsection{Can the second law be violated with P-CTC?}

A P-CTC model has a distinctive feature -- one can post-select with certainty on any measurement outcome \cite{Lloyd2011PRL,Lloyd2011PRD}, even using a single qubit \cite{BrunWilde2011}. Since the output state of a P-CTC circuit can be chosen with certainty, depending on the initial state one should be able to either increase or decrease entropy. It remains a matter of accurate unitary operations selection to achieve either case. 

In the next subsection we intentionally aim to violate the second law of thermodynamics. To this end we construct a P-CTC-assisted circuit with an appropriately chosen unitary operations which lead to decrease of entropy. In particular, we transform an initial ensemble of pure states into a single pure state of choice, e.g. $|0\rangle$.

For clarity, we follow the analogical approaches as in the case of D-CTC model, i.e., first we evolve the components of the initial proper mixture and then average over evolutions; in the second approach we evolve the whole mixture. Note that in our example, one approach seems sufficient because regardless of weights in the final mixture  (either unaffected or updated), the resulting output state still will be $|0\rangle$.

\subsection{Approach 1: Violation of the second law of thermodynamics in a P-CTC model}\label{PCTCapproach1}
Consider state of the first qudit:
\be\label{pctcstate1}
\rho_1=\sum_l p_l |\psi_l\rangle_1 \langle \psi_l|_1
\ee
which follows from a statistically prepared ensemble of initial states:
\be\label{pctcpurestates}
|\psi_l\rangle_1 = \sum_{i=0}^{d-1} a_{li} |i\rangle_1
\ee
that satisfy the normalization condition
\be\label{pctcnormalization}
\sum_{i=0}^{d-1} |a_{li}|^2=1.
\ee
The maximally entangled state of the second and the third qudit is given by:
\be\label{pctcentangled}
|\Phi^+\rangle_{23} = \frac{1}{\sqrt{d}} \sum_{j=0}^{d-1} |j\rangle_2 \ot |j\rangle_3.
\ee
Since we first perform evolution on each component of an initial mixture separately, we begin with a pure state of a whole system:
\be
\label{initialstate123}
|\psi\rangle_{123}\!=\!|\psi_l\rangle_1 \!\ot\! |\Phi^+\rangle_{23}\!=\!\frac{1}{\sqrt{d}} \sum_{i,j=0}^{d-1} a_{li} |i\rangle_1 \ot |j\rangle_2 \ot |j\rangle_3,
\ee
where we used Eqs. \eqref{pctcpurestates} and \eqref{pctcentangled}.

As showed in Fig. \ref{ctcrho1i2} b), the first two qudits evolve due to the controlled-unitary operation $U_{12}$, whereas the third qudit remains untouched.  We choose the overall evolution $U$:
\be
\label{unitary}
U=U_{12} \ot I_3=( \sum_{i=0}^{d-1} |i\rangle \langle i| \ot U_{i})_{12} \ot I_3
\ee
such that 
\be\label{pctcUi}
U_{i}=\sum_{j=0}^{d-1} |f_i(j)\rangle_2 \langle j|_2,
\ee
where $f_i(j)$ is a function to be chosen later on. The initial state of the system \eqref{initialstate123} evolves then as:
\be
U|\psi\rangle_{123}=\frac{1}{\sqrt{d}} \sum_{i,j=0}^{d-1} a_{li} |i\rangle_1 \ot |f_i(j)\rangle_2 \ot  |j\rangle_3.
\ee
Next step is to post-select the state of second qudit with certainty. After renormalization it reads as:
\be
\label{secondqudit}
|\psi\rangle_{2}'\!\!=\!\!\frac{\langle \Phi^+|_{13} \left(U|\psi\rangle_{123}\right)}{|\langle \Phi^+|_{13} \left(U|\psi\rangle_{123}\right)|}\!=\!\!\frac{\sum_{i=0}^{d-1} a_{li} |f_i(i)\rangle_2}{|\sum_{i,k=0}^{d-1} a_{li} a_{lk}^*\langle f_k(k)|f_i(i)\rangle_2|}.
\ee

Let us now impose the following two conditions. First, as established earlier, we choose the post-measurement state to be $|\psi\rangle_{2}'=|0\rangle$, which requires that
\be
\label{condition1}
f_i(i)=0, \quad \forall_i.
\ee
Second, the operation $U$ must be unitary, i.e., $UU^{\dagger}=U^{\dagger}U=1$, which simplifies to 
\be
\label{condition2}
\quad U_i U_i^\dagger = U_i^\dagger U_i  =1, \quad \forall_i.
\ee

To fulfill the first condition \eqref{condition1}, we choose functions $f_i(j)$ as permutation functions shown graphically in Fig. \ref{permutation} and defined explicitly in the Appendix \ref{APPENDIXpermutations}. Indeed $f_i(i)=0, \forall_i$, and so the condition \eqref{condition1} is obeyed. Moreover, since each $f_i(j)$ represents permutation, the second condition \eqref{condition2} is immediately satisfied.

Eventually, using a condition \eqref{condition1} in Eq. \eqref{secondqudit}, one sees that the P-CTC-assisted circuit can transform each component of a mixture into the same state (here: $|0\rangle$). After averaging over evolutions, we obtain that the initially mixed state becomes a pure state with entropy zero. Consequently, we can observe the decrease in entropy and that the second law of thermodynamics is violated.

\begin{figure}[t]
\centering
\includegraphics[width=0.48\textwidth]{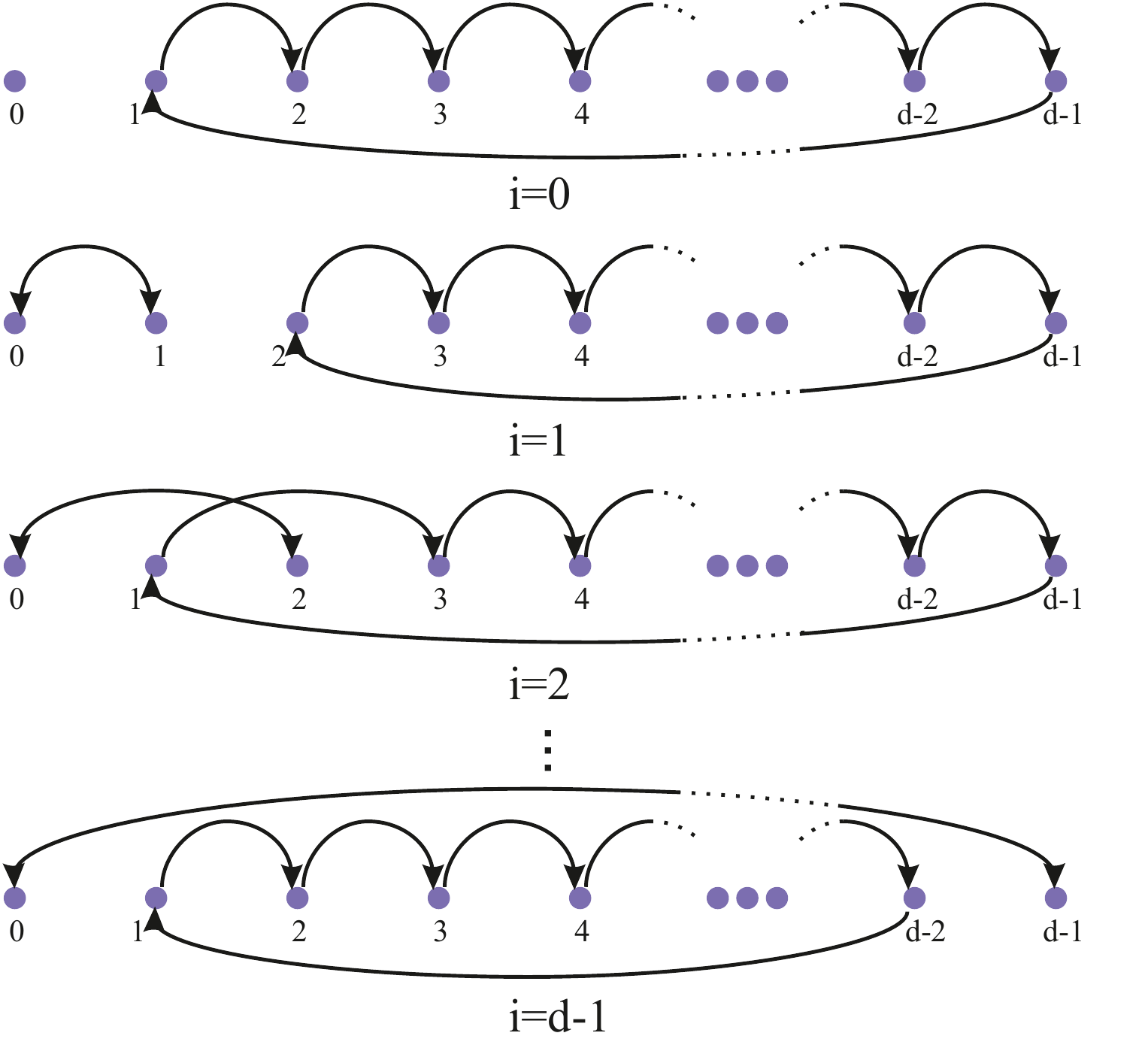}
\caption{Illustration of the action of permutation functions $f_i(j)$ for $i,j=\{0,1,..,d-1\}$. The special cases for $i=\{0,1,d-1\}$ are explicitly depicted, whereas the omitted illustrations for the cases for $i\geq 3$ are analogous as for $i=2$.}
\label{permutation}
\end{figure}

\subsection{Approach 2: Violation of the second law of thermodynamics in a P-CTC model}\label{PCTCapproach2}
In the second approach we evolve the full mixture (cf. Eq. \eqref{statepctc}):
\ben
\label{initialrho}
\rho=\frac1d \sum_{l=0}^{N} p_l \!\!\! \sum_{\substack{m,n,\\ k,j=0}}^{d-1} \!\!\! a_{lk} a^{*}_{lj} |k\rangle_{1}  \langle j|_{1} \!\!\ot \!|m\rangle_{2}  \langle n|_{2} \!\ot\! |m\rangle_{3}  \langle n|_{3},
\een
where we used Eqs. \eqref{pctcstate1}, \eqref{pctcpurestates} and \eqref{pctcentangled}, and where coefficients satisfy the normalization condition
\be\label{pctcnormalization2}
\sum_{k=0}^{d-1} |a_{lk}|^2=1.
 \ee
First, the above mixture \eqref{initialrho} undergoes the unitary evolution given by operator $U$ defined in Eq. \eqref{unitary} with \eqref{pctcUi}. Next, a transformed state of a whole system $\rho'$ is projected into an entangled state $|\Phi^+\rangle_{13}$ with probability 1. This gives the final unnormalized state of qudit 2 in the form (cf. \eqref{projectionpctc}):
\ben
\nonumber
\tilde{\rho}_2' &=& \< \Phi^+|_{13} U\rho  U^{\dagger} |\Phi^+\>_{13}\\
&=& \frac{1}{d^2} \sum_{l=0}^{N} p_l  \sum_{k,j=0}^{d-1}  a_{lk} a^{*}_{lj} |f_k(k)\>_2 \<f_j(j)|_2.
\een
Following the condition  \eqref{condition1}, we obtain that $\forall _k f_k(k)=0$ and $\forall _j f_j(j)=0$, and therefore the state \eqref{projectionpctc} is a mixture of pure states $|0\>$. Although the upcoming renormalization would alter the original weights $p_l$ in the mixture, it does not affect the result since we mix only one pure state $|0\>$ with different weights. Eventually, we obtain a final pure state: $\rho'= |0\>_2 \<0|_2$ with entropy zero.

Again, in this approach, we observe that the P-CTC-assisted circuit with appropriately satisfied conditions \eqref{condition1} and \eqref{condition2}, allows to transform a mixture of states into a state with zero entropy. If the entropy of initial state is nonzero, we observe the decrease of entropy throughout the procedure and therefore violation of the second law.

\section{DISCUSSION}
In our work we have examined whether the second law of thermodynamics holds under the two basic models of CTC-assisted quantum circuits: Deutsch's (D-CTC) and post-selected (P-CTC). For a D-CTC model we established that if we let ourselves ``fall into a linearity trap" by calculating the average of evolutions performed individually on the components of an initial mixture, then the total entropy of the system decreases and therefore the second law of thermodynamics is violated. On the other hand, when we avoid falling into such a trap, i.e., the evolution is performed on the whole input state, the total entropy does not decrease. In the P-CTC model, however, regardless of approach, the second law of thermodynamics is violated.

There are two ways in which one can interpret the obtained results for a D-CTC model. Firstly, one can look at it in such a way: the system whose entropy we study moves forward in time. In this case, one could expect the second law of thermodynamics to hold, especially since the entropy is calculated at the points in time where CTC either does not yet or no longer exists. Therefore, if in this case our approach 1, i.e.,``falling into a linearity trap", allows to violate the second law, one could conclude that either of two things can happen: a) the approach is incorrect or b) a D-CTC model itself is inaccurate.

On the other hand, one may look at the problem from a very different perspective. Namely that a quantum system, whose entropy is studied, interacts for a finite period of time with some other system which travels backwards in time. This may influence the entropy in yet unknown manner. In this case, we cannot definitively state what to expect, but the violation of the second law could remain a possibility. Therefore, our conclusions from the second approach of ``avoiding linearity trap" which indicate that entropy always increases (and so the second law always holds) could be a surprising result that casts doubt on a D-CTC model.

A separate issue is the matter of whether the approaches of ``falling into" or ``avoiding a linearity trap" are correct at all. In literature \cite{Bennett2009,Ahn2010,BrunWinter2013,Moulick2016,BrunWilde2009,Brun2003,Bacon2004,Aaronson2009,RalphMyers2010,Cavalcanti2010}, the peculiar implications of the CTC-based models were always related to the chosen approach. If the equivalent of our first approach was chosen, then peculiar phenomena such as perfect cloning or distinguishing of nonorthogonal states were observed. Likewise, we observed something bizzare -- a violation of the second law of thermodynamics. Only by using Bennett's \textit{et al.} argument we could ``save" the second law, just as Bennett \textit{et al.} have ``saved" the quantum mechanics laws in the presence of CTC.  

In the case of P-CTC model, both of our approaches led to the same conclusion: the second law of thermodynamics was violated. This was however only a result of our choice of unitary operations and the final state that was obtained upon their use. A P-CTC model is based on the post-selection with certainty. Using this feature, we performed the unitary operation in such a way that the output was a state with lower entropy, and we showed that it is indeed achievable. However, equally likely one could choose such operations that would lead to the entropy either greater or equal to the entropy of initial state. 

Now, we encounter the same issue as previously for the D-CTC model.  It involves one of the fundamental long-standing open problems in physics i.e., the nature of time itself and its role in the description of physical reality including CTCs. In particular, we see that if our thermodynamic criterion would imply so called ``thermodynamic arrow of time" \cite{Monsel2018,Batalh2015}, which seems to``tolerate" D-CTCs more than P-CTCs, then the D-CTC model could turn out to be more reasonable.

It is impossible to unambiguously determine this dispute in favour of any of the aforementioned models of CTCs. However, the importance of the investigations performed here is highlighted by the fact that now the characteristics of these models can be better understood and provide an indispensable basis for future investigations. Moreover, we have shown that careful consideration of the approach used in CTC-assisted quantum circuits may lead to different implications on other related theories, such as thermodynamics. Remarkably the latter has been recently considered \cite{Bera2017} as a consequence of information conservation \cite{Horodecki2005, Hulpke2006}. Then it cannot be excluded that the information conservation principle is a necessary condition for physical (nonlinear) extensions of the quantum theory.

Of course, other issues remain as to what are the physical meanings of the Deutsch's type models  and whether their physical implications can be truly observed using experimental methodologies. Finally, the investigation of such models may also shed some light on the epistemic and formal gap between general relativity and quantum mechanics \cite{Greene2004, KraussWilczek2014}.

\begin{acknowledgments}
We would like to thank Charles Bennett, Micha\l{} Horodecki, Waldemar K\l{}obus, Debbie Leung, and Subhayan Roy Moulick for interesting and helpful discussions. The authors also thank the anonymous reviewers for constructive comments and suggestions. This work was supported by National Science Centre, Poland, grant OPUS 9. 2015/17/B/ST2/01945 and by John Templeton Foundation through the  Grant ID $\#56033$.  The  opinions expressed in this publication are those of the authors and do not necessarily reflect the views of the John Templeton Foundation.
\end{acknowledgments}

\onecolumngrid
\begin{appendix}
\section{ Proof of the self-consistency condition}\label{APPENDIXselfcons}
Here we show that $\rho_{CTC}=|k\>\<k|$ constitutes a unique solution of a self-consistent condition \eqref{selfcons}. The state transformation is specified by the unitary operation $U_{SCTC}$ introduced in the main text in Eqs. \eqref{unitaryDeutsch} and \eqref{opU} -- \eqref{opV}. The initial pure state of a system S that, together with a state of CTC system, undergoes the above transformation is given by $|\psi_k\>$.

We begin by showing that $\rho_{CTC}=|k\>\<k|$ satisfies the self-consistent condition \eqref{selfcons}. RHS of \eqref{selfcons} with  $\rho_{CTC}=|k\>\<k|$ reads as:
\ben
\nonumber
\!\!&\Tr_{S}& \left(\!\! \left(\sum_{i} V_{i} \otimes |i\rangle \langle i| \right) \!\! \left(\sum_{i'}|i'\rangle \langle i'| \otimes U_{i'} \right)\!\! \text{SWAP}\left(|\psi_{k}\rangle \langle \psi_{k}|\otimes |k\rangle \langle k|  \right) \text{SWAP}\!\!\left(\sum_{j'}|j'\rangle \langle j'| \otimes U_{j'}^{\dagger} \right)\!\! \left(\sum_{j} V_{j}^{\dagger} \otimes |j\rangle \langle j|\right)\!\! \right)\\
\nonumber
&=&\!\!\Tr_{S} \left(\!\! \left(\sum_{i} V_{i} \otimes |i\rangle \langle i| \right) \!\! \left(\sum_{i'}|i'\rangle \langle i'| \otimes U_{i'} \right)\!\!\left(|k\rangle \langle k| \otimes |\psi_{k}\rangle \langle \psi_{k}| \right)\!\!\left(\sum_{j'}|j'\rangle \langle j'| \otimes U_{j'}^{\dagger} \right)\!\! \left(\sum_{j} V_{j}^{\dagger} \otimes |j\rangle \langle j|\right)\!\! \right)\\
\nonumber
&=&\!\!\Tr_{S}\left( \left( \sum_{i} V_{i} \otimes |i\rangle \langle i| \right)  \left( |k\rangle \langle k|  \otimes U_{k} |\psi_{k}\rangle \langle \psi_{k}| U_{k}^{\dagger} \right) \left( \sum_{j} V_{j}^{\dagger} \otimes |j\rangle \langle j| \right) \right)\\
\nonumber
&=&\!\!\Tr_{S}\left( \left( \sum_{i} V_{i} \otimes |i\rangle \langle i| \right)  \left( |k\rangle \langle k|  \otimes |k\rangle \langle k| \right) \left( \sum_{j} V_{j}^{\dagger} \otimes |j\rangle \langle j| \right) \right)\\
\nonumber
&=&\!\!\Tr_{S} \left( V_{k} |k\rangle \langle k| V_{k}^{\dagger} \otimes |k\rangle \langle k| \right)\\
\nonumber
&=&\!\!\Tr_{S} \left(|\varphi\rangle \langle\varphi| \otimes |k\rangle \langle k| \right)\\
&=&\!\!|k\rangle \langle k| \equiv \rho_{CTC}.
\een
In the third equality we used Eq.  \eqref{opU} and in the fifth equation Eq. \eqref{opV}.

Next, we prove that this solution is unique. For this reason, let us consider the self-consistent condition \eqref{selfcons} once again, but now for a general state of a CTC system in the form $\rho_{CTC}=\sum_{m,n} \rho_{mn} |m\rangle \langle n|$. To show the uniqueness of our solution, we need to prove that all coefficients $\rho_{mn}$ are null, except for one which is equal to one: $\rho_{kk}=1$. To this end, we simplify the self-consistency condition \eqref{selfcons} as follows:

\ben
\label{proofunique}
\nonumber
\rho_{CTC}\!\! &=&\!\!\Tr_{S} \left(\!\! \left(\sum_{i} V_{i} \otimes |i\rangle \langle i| \right) \!\! \left(\sum_{i'}|i'\rangle \langle i'| \otimes U_{i'} \right)\!\! \left(\sum_{m,n} \rho_{mn} |m\rangle \langle n| \otimes |\psi_{k}\rangle \langle \psi_{k}| \right) \!\!\left(\sum_{j'}|j'\rangle \langle j'| \otimes U_{j'}^{\dagger} \right)\!\! \left(\sum_{j} V_{j}^{\dagger} \otimes |j\rangle \langle j|\right)\!\! \right)\\
\nonumber
&=&\!\!\Tr_{S}\left( \left( \sum_{i} V_{i} \otimes |i\rangle \langle i| \right)  \left(\sum_{m,n} \rho_{mn} |m\rangle \langle n| \otimes U_{m} |\psi_{k}\rangle \langle \psi_{k}| U_{n}^{\dagger} \right) \left( \sum_{j} V_{j}^{\dagger} \otimes |j\rangle \langle j| \right) \right)\\
\nonumber
&=&\!\!\Tr_{S} \left( \sum_{i,j,m,n} \rho_{mn} V_{i} |m\rangle \langle n| V_{j}^{\dagger} \otimes |i\rangle \langle i| U_{m} |\psi_k\rangle \langle \psi_k| U_n^{\dagger} |j\rangle \langle j| \right)\\
\nonumber
&=&\!\!\sum_{i,j,l,m,n} \rho_{mn} \langle l| V_i |m\rangle \langle n| V_j^{\dagger} |l\rangle |i\rangle \langle i| U_{m} |\psi_k\rangle \langle \psi_k| U_n^{\dagger} |j\rangle \langle j|\\
&=&\!\!\sum_{i,j} \left(\sum_{m,n}  \rho_{mn} \langle n| V_j^{\dagger} V_i |m\rangle \langle i| U_m |\psi_k\rangle \langle \psi_k| U_n^{\dagger} |j\rangle \right) |i\rangle  \langle j|.
\een
Note that in the first line, the SWAP operation has been already executed. Now, let us write the LHS of  Eq. \eqref{proofunique} as 
\be\label{densitymatrix}
\rho_{CTC}=\sum_{i,j} \rho_{ij} |i\rangle \langle j|.
\ee
From Eqs. \eqref{proofunique} and \eqref{densitymatrix} we get that
\be
\label{rhoij}
\rho_{ij}=\sum_{m,n} \rho_{mn} \langle n| V_j^{\dagger} V_i |m\rangle \langle i| U_m |\psi_k\rangle \langle \psi_k| U_n^{\dagger} |j\rangle.
\ee

As mentioned earlier, to prove the uniqueness of the solution $\rho_{CTC}=|k\> \<k|$, we need to show that only one coefficient from $\rho_{ij}$ \eqref{rhoij} is nonzero and equals to one: $\rho_{kk}=1$. Let us then rewrite Eq. \eqref{rhoij} by substituting $i=j=k$, and simplifying further:
\ben
\label{rhokk}
\nonumber
\rho_{kk}&=&\sum_{m,n} \rho_{mn} \langle n| V_k^{\dagger} V_k |m\rangle \langle k| U_m |\psi_k\rangle \langle \psi_k| U_n^{\dagger} |k\rangle\\
\nonumber
&=&\sum_{m} \rho_{mm} |\langle k| U_m |\psi_k\rangle|^2\\
\nonumber
&=&\rho_{kk} |\langle k| U_k |\psi_k\rangle|^2 + \sum_{m \neq k} \rho_{mm} |\langle k| U_m |\psi_k\rangle|^2\\
&=& \rho_{kk}  + \sum_{m \neq k} \rho_{mm} |\langle k| U_m |\psi_k\rangle|^2.
\een
Given that the operators $\{U_k \}$, $\{V_k\}$ form sets of unitary transformations, in the second equality we used that fact that $V_k^{\dagger} V_k=I$. Next, we splitted the RHS of  \eqref{rhokk} into two expressions: first for $m=k$, and second for the remaining $m$'s. In the last equality we used the relation  \eqref{opU} from the main text.

At this point we got to the Eq. (6) from the proof of Brun \textit{et.al} \cite{BrunWilde2009}. The authors proceed there by formulating two sufficient conditions for the solution to be unique. First, $U_{m} |\psi_{m}\> = |m\>, \forall_m$, that corresponds to our Eq. \eqref{opU}, and second, $\langle k| U_m |\psi_k\rangle \neq 0, \forall_{k,m}$ which implies $\rho_{mm}=0$ for all $m\neq k$, and therefore $\rho_{kk}=1$. Authors complete the proof by showing the construction of a set of unitary operators $\{U_m\}$ that satisfy the above conditions. The rest of our proof follows the same steps.

\section{Explicit form of the permutation functions $f_i(j)$}\label{APPENDIXpermutations}
Below we provide an explicit form of the permutation functions $f_i(j)$ represented graphically in the main text in Fig. 3.
The functions $f_i(j)$ are given by:
\be
 f_{i \neq \{1,d-1\}} (j)=
  \begin{cases}
    j \rightarrow  j+ 1, \quad &\forall_{j \neq 0,i,i-1,d-1}\\
    i \leftrightarrow  0,\\
     (i - 1) \text{mod } d \rightarrow   i+1,\\
    d-1 \rightarrow   1, 
  \end{cases}
\label{functioni}
\ee
and for special cases of $i=1$ and $i=d-1$, respectively by:
\be
 f_1 (j)=
  \begin{cases}
    j \rightarrow  j+ 1, \quad &\forall_{j \neq 0,1,d-1}\\
    1 \leftrightarrow  0,\\
    d-1 \rightarrow   2, 
  \end{cases}
\label{function1}
\ee
and
\be
 f_{d-1} (j)=
  \begin{cases}
    j \rightarrow  j+ 1, \quad &\forall_{j \neq 0,d-2,d-1}\\
    d-1 \leftrightarrow  0,\\
    d-2 \rightarrow   1.
  \end{cases}
\label{functiond1}
\ee
Such definition assures that the operation $U$ \eqref{unitary}--\eqref{pctcUi}, defined in terms of $f_i(j)$, is unitary, and also that $f_i(i)=0$ for every $i$, as demanded.
\end{appendix}


\begin{thebibliography}{34}%
\makeatletter
\providecommand \@ifxundefined [1]{%
 \@ifx{#1\undefined}
}%
\providecommand \@ifnum [1]{%
 \ifnum #1\expandafter \@firstoftwo
 \else \expandafter \@secondoftwo
 \fi
}%
\providecommand \@ifx [1]{%
 \ifx #1\expandafter \@firstoftwo
 \else \expandafter \@secondoftwo
 \fi
}%
\providecommand \natexlab [1]{#1}%
\providecommand \enquote  [1]{``#1''}%
\providecommand \bibnamefont  [1]{#1}%
\providecommand \bibfnamefont [1]{#1}%
\providecommand \citenamefont [1]{#1}%
\providecommand \href@noop [0]{\@secondoftwo}%
\providecommand \href [0]{\begingroup \@sanitize@url \@href}%
\providecommand \@href[1]{\@@startlink{#1}\@@href}%
\providecommand \@@href[1]{\endgroup#1\@@endlink}%
\providecommand \@sanitize@url [0]{\catcode `\\12\catcode `\$12\catcode
  `\&12\catcode `\#12\catcode `\^12\catcode `\_12\catcode `\%12\relax}%
\providecommand \@@startlink[1]{}%
\providecommand \@@endlink[0]{}%
\providecommand \url  [0]{\begingroup\@sanitize@url \@url }%
\providecommand \@url [1]{\endgroup\@href {#1}{\urlprefix }}%
\providecommand \urlprefix  [0]{URL }%
\providecommand \Eprint [0]{\href }%
\providecommand \doibase [0]{http://dx.doi.org/}%
\providecommand \selectlanguage [0]{\@gobble}%
\providecommand \bibinfo  [0]{\@secondoftwo}%
\providecommand \bibfield  [0]{\@secondoftwo}%
\providecommand \translation [1]{[#1]}%
\providecommand \BibitemOpen [0]{}%
\providecommand \bibitemStop [0]{}%
\providecommand \bibitemNoStop [0]{.\EOS\space}%
\providecommand \EOS [0]{\spacefactor3000\relax}%
\providecommand \BibitemShut  [1]{\csname bibitem#1\endcsname}%
\let\auto@bib@innerbib\@empty
\bibitem [{\citenamefont {G\"odel}(1949)}]{Godel1949}%
  \BibitemOpen
  \bibfield  {author} {\bibinfo {author} {\bibfnamefont {K.}~\bibnamefont
  {G\"odel}},\ }\href {\doibase 10.1103/RevModPhys.21.447} {\bibfield
  {journal} {\bibinfo  {journal} {Rev. Mod. Phys.}\ }\textbf {\bibinfo {volume}
  {21}},\ \bibinfo {pages} {447} (\bibinfo {year} {1949})}\BibitemShut
  {NoStop}%
\bibitem [{\citenamefont {Morris}\ \emph {et~al.}(1988)\citenamefont {Morris},
  \citenamefont {Thorne},\ and\ \citenamefont {Yurtsever}}]{Thorne1988}%
  \BibitemOpen
  \bibfield  {author} {\bibinfo {author} {\bibfnamefont {M.~S.}\ \bibnamefont
  {Morris}}, \bibinfo {author} {\bibfnamefont {K.~S.}\ \bibnamefont {Thorne}},
  \ and\ \bibinfo {author} {\bibfnamefont {U.}~\bibnamefont {Yurtsever}},\
  }\href {\doibase 10.1103/PhysRevLett.61.1446} {\bibfield  {journal} {\bibinfo
   {journal} {Phys. Rev. Lett.}\ }\textbf {\bibinfo {volume} {61}},\ \bibinfo
  {pages} {1446} (\bibinfo {year} {1988})}\BibitemShut {NoStop}%
\bibitem [{\citenamefont {Nahin}(1999)}]{Nahin1999}%
  \BibitemOpen
  \bibfield  {author} {\bibinfo {author} {\bibfnamefont {P.~J.}\ \bibnamefont
  {Nahin}},\ }\href@noop {} {\emph {\bibinfo {title} {Time Machines: Time
  Travel in Physics, Metaphysics and Science Fiction}}}\ (\bibinfo  {publisher}
  {Springer-Verlag and AIP Press, New York},\ \bibinfo {year}
  {1999})\BibitemShut {NoStop}%
\bibitem [{\citenamefont {Deutsch}(1991)}]{Deutsch1991}%
  \BibitemOpen
  \bibfield  {author} {\bibinfo {author} {\bibfnamefont {D.}~\bibnamefont
  {Deutsch}},\ }\href {\doibase 10.1103/PhysRevD.44.3197} {\bibfield  {journal}
  {\bibinfo  {journal} {Phys. Rev. D}\ }\textbf {\bibinfo {volume} {44}},\
  \bibinfo {pages} {3197} (\bibinfo {year} {1991})}\BibitemShut {NoStop}%
\bibitem [{\citenamefont {Friedman}\ \emph {et~al.}(1990)\citenamefont
  {Friedman}, \citenamefont {Morris}, \citenamefont {Novikov}, \citenamefont
  {Echeverria}, \citenamefont {Klinkhammer}, \citenamefont {Thorne},\ and\
  \citenamefont {Yurtsever}}]{Novikov1990}%
  \BibitemOpen
  \bibfield  {author} {\bibinfo {author} {\bibfnamefont {J.}~\bibnamefont
  {Friedman}}, \bibinfo {author} {\bibfnamefont {M.~S.}\ \bibnamefont
  {Morris}}, \bibinfo {author} {\bibfnamefont {I.~D.}\ \bibnamefont {Novikov}},
  \bibinfo {author} {\bibfnamefont {F.}~\bibnamefont {Echeverria}}, \bibinfo
  {author} {\bibfnamefont {G.}~\bibnamefont {Klinkhammer}}, \bibinfo {author}
  {\bibfnamefont {K.~S.}\ \bibnamefont {Thorne}}, \ and\ \bibinfo {author}
  {\bibfnamefont {U.}~\bibnamefont {Yurtsever}},\ }\href {\doibase
  10.1103/PhysRevD.42.1915} {\bibfield  {journal} {\bibinfo  {journal} {Phys.
  Rev. D}\ }\textbf {\bibinfo {volume} {42}},\ \bibinfo {pages} {1915}
  (\bibinfo {year} {1990})}\BibitemShut {NoStop}%
\bibitem [{\citenamefont {Ahn}\ \emph {et~al.}(2010)\citenamefont {Ahn},
  \citenamefont {Ralph},\ and\ \citenamefont {Mann}}]{Ahn2010}%
  \BibitemOpen
  \bibfield  {author} {\bibinfo {author} {\bibfnamefont {D.}~\bibnamefont
  {Ahn}}, \bibinfo {author} {\bibfnamefont {T.~C.}\ \bibnamefont {Ralph}}, \
  and\ \bibinfo {author} {\bibfnamefont {R.~B.}\ \bibnamefont {Mann}},\
  }\href@noop {} {\  (\bibinfo {year} {2010})},\ \Eprint
  {http://arxiv.org/abs/1008.0221} {arXiv:1008.0221 [quant-ph]} \BibitemShut
  {NoStop}%
\bibitem [{\citenamefont {Brun}\ \emph {et~al.}(2013)\citenamefont {Brun},
  \citenamefont {Wilde},\ and\ \citenamefont {Winter}}]{BrunWinter2013}%
  \BibitemOpen
  \bibfield  {author} {\bibinfo {author} {\bibfnamefont {T.~A.}\ \bibnamefont
  {Brun}}, \bibinfo {author} {\bibfnamefont {M.~M.}\ \bibnamefont {Wilde}}, \
  and\ \bibinfo {author} {\bibfnamefont {A.}~\bibnamefont {Winter}},\ }\href
  {\doibase 10.1103/PhysRevLett.111.190401} {\bibfield  {journal} {\bibinfo
  {journal} {Phys. Rev. Lett.}\ }\textbf {\bibinfo {volume} {111}},\ \bibinfo
  {pages} {190401} (\bibinfo {year} {2013})}\BibitemShut {NoStop}%
\bibitem [{\citenamefont {Moulick}\ and\ \citenamefont
  {Panigrahi}(2016)}]{Moulick2016}%
  \BibitemOpen
  \bibfield  {author} {\bibinfo {author} {\bibfnamefont {S.~R.}\ \bibnamefont
  {Moulick}}\ and\ \bibinfo {author} {\bibfnamefont {P.~K.}\ \bibnamefont
  {Panigrahi}},\ }\href {\doibase doi:10.1038/srep37958} {\bibfield  {journal}
  {\bibinfo  {journal} {Sci. Rep.}\ }\textbf {\bibinfo {volume} {6}},\ \bibinfo
  {pages} {37958} (\bibinfo {year} {2016})}\BibitemShut {NoStop}%
\bibitem [{\citenamefont {Brun}\ \emph {et~al.}(2009)\citenamefont {Brun},
  \citenamefont {Harrington},\ and\ \citenamefont {Wilde}}]{BrunWilde2009}%
  \BibitemOpen
  \bibfield  {author} {\bibinfo {author} {\bibfnamefont {T.~A.}\ \bibnamefont
  {Brun}}, \bibinfo {author} {\bibfnamefont {J.}~\bibnamefont {Harrington}}, \
  and\ \bibinfo {author} {\bibfnamefont {M.~M.}\ \bibnamefont {Wilde}},\ }\href
  {\doibase 10.1103/PhysRevLett.102.210402} {\bibfield  {journal} {\bibinfo
  {journal} {Phys. Rev. Lett.}\ }\textbf {\bibinfo {volume} {102}},\ \bibinfo
  {eid} {210402} (\bibinfo {year} {2009})}\BibitemShut {NoStop}%
\bibitem [{\citenamefont {Ringbauer}\ \emph {et~al.}(2014)\citenamefont
  {Ringbauer}, \citenamefont {Broome}, \citenamefont {Myers}, \citenamefont
  {White},\ and\ \citenamefont {Ralph}}]{Ralph2014}%
  \BibitemOpen
  \bibfield  {author} {\bibinfo {author} {\bibfnamefont {M.}~\bibnamefont
  {Ringbauer}}, \bibinfo {author} {\bibfnamefont {M.~A.}\ \bibnamefont
  {Broome}}, \bibinfo {author} {\bibfnamefont {C.~R.}\ \bibnamefont {Myers}},
  \bibinfo {author} {\bibfnamefont {A.~G.}\ \bibnamefont {White}}, \ and\
  \bibinfo {author} {\bibfnamefont {T.~C.}\ \bibnamefont {Ralph}},\ }\href
  {\doibase 10.1038/ncomms5145} {\bibfield  {journal} {\bibinfo  {journal}
  {Nat. Comm.}\ }\textbf {\bibinfo {volume} {4145}} (\bibinfo {year} {2014})}
  \BibitemShut {NoStop}%
\bibitem [{\citenamefont {Brun}(2003)}]{Brun2003}%
  \BibitemOpen
  \bibfield  {author} {\bibinfo {author} {\bibfnamefont {T.~A.}\ \bibnamefont
  {Brun}},\ }\href {\doibase 10.1023/A:1025967225931} {\bibfield  {journal}
  {\bibinfo  {journal} {Found. Phys.}\ }\textbf {\bibinfo {volume} {16}},\
  \bibinfo {pages} {245} (\bibinfo {year} {2003})}\BibitemShut {NoStop}%
\bibitem [{\citenamefont {Bacon}(2004)}]{Bacon2004}%
  \BibitemOpen
  \bibfield  {author} {\bibinfo {author} {\bibfnamefont {D.}~\bibnamefont
  {Bacon}},\ }\href {\doibase 10.1103/PhysRevA.70.032309} {\bibfield  {journal}
  {\bibinfo  {journal} {Phys. Rev. A}\ }\textbf {\bibinfo {volume} {70}},\
  \bibinfo {pages} {032309} (\bibinfo {year} {2004})}\BibitemShut {NoStop}%
\bibitem [{\citenamefont {Aaronson}\ and\ \citenamefont
  {Watrous}(2009)}]{Aaronson2009}%
  \BibitemOpen
  \bibfield  {author} {\bibinfo {author} {\bibfnamefont {S.}~\bibnamefont
  {Aaronson}}\ and\ \bibinfo {author} {\bibfnamefont {J.}~\bibnamefont
  {Watrous}},\ }\href {\doibase 10.1098/rspa.2008.0350} {\bibfield  {journal}
  {\bibinfo  {journal} {Proc. R. Soc. A}\ }\textbf {\bibinfo {volume} {465}},\ \bibinfo {pages}
  {631} (\bibinfo {year} {2009})}\BibitemShut {NoStop}%
\bibitem [{\citenamefont {{Bennett}}\ \emph {et~al.}(2009)\citenamefont
  {{Bennett}}, \citenamefont {{Leung}}, \citenamefont {{Smith}},\ and\
  \citenamefont {{Smolin}}}]{Bennett2009}%
  \BibitemOpen
  \bibfield  {author} {\bibinfo {author} {\bibfnamefont {C.~H.}\ \bibnamefont
  {{Bennett}}}, \bibinfo {author} {\bibfnamefont {D.}~\bibnamefont {{Leung}}},
  \bibinfo {author} {\bibfnamefont {G.}~\bibnamefont {{Smith}}}, \ and\
  \bibinfo {author} {\bibfnamefont {J.~A.}\ \bibnamefont {{Smolin}}},\
  }\href {https://link.aps.org/doi/10.1103/PhysRevLett.103.170502} {\bibfield  {journal} {\bibinfo  {journal} {Phys. Rev. Lett.}\
  }\textbf {\bibinfo {volume} {103}},\ \bibinfo {eid} {170502} (\bibinfo {year}
  {2009})} \BibitemShut {NoStop}%
\bibitem [{\citenamefont {{Bennett}}(2005)}]{Bennett2005}%
  \BibitemOpen
  \bibfield  {author} {\bibinfo {author} {\bibfnamefont {C.~H.}\ \bibnamefont
  {{Bennett}}},\ }\href@noop {} {\bibfield  {journal} {\bibinfo  {journal}
  {QUPON talk}\ } (\bibinfo {year} {2005})},\ \bibinfo {note}
  {\href{http://web.archive.org/web/20070206131550/http://www.research.ibm.com/people/b/bennetc/QUPONBshort.pdf}{http://web.archive.org/web/20070206131550/http://
  www.research.ibm.com/people/b/bennetc/QUPONB short.pdf}}\BibitemShut
  {NoStop}%
\bibitem [{\citenamefont {Lloyd}\ \emph
  {et~al.}(2011{\natexlab{a}})\citenamefont {Lloyd}, \citenamefont {Maccone},
  \citenamefont {Garcia-Patron}, \citenamefont {Giovannetti}, \citenamefont
  {Shikano}, \citenamefont {Pirandola}, \citenamefont {Rozema}, \citenamefont
  {Darabi}, \citenamefont {Soudagar}, \citenamefont {Shalm},\ and\
  \citenamefont {Steinberg}}]{Lloyd2011PRL}%
  \BibitemOpen
  \bibfield  {author} {\bibinfo {author} {\bibfnamefont {S.}~\bibnamefont
  {Lloyd}}, \bibinfo {author} {\bibfnamefont {L.}~\bibnamefont {Maccone}},
  \bibinfo {author} {\bibfnamefont {R.}~\bibnamefont {Garcia-Patron}}, \bibinfo
  {author} {\bibfnamefont {V.}~\bibnamefont {Giovannetti}}, \bibinfo {author}
  {\bibfnamefont {Y.}~\bibnamefont {Shikano}}, \bibinfo {author} {\bibfnamefont
  {S.}~\bibnamefont {Pirandola}}, \bibinfo {author} {\bibfnamefont {L.~A.}\
  \bibnamefont {Rozema}}, \bibinfo {author} {\bibfnamefont {A.}~\bibnamefont
  {Darabi}}, \bibinfo {author} {\bibfnamefont {Y.}~\bibnamefont {Soudagar}},
  \bibinfo {author} {\bibfnamefont {L.~K.}\ \bibnamefont {Shalm}}, \ and\
  \bibinfo {author} {\bibfnamefont {A.~M.}\ \bibnamefont {Steinberg}},\ }\href
  {\doibase 10.1103/PhysRevLett.106.040403} {\bibfield  {journal} {\bibinfo
  {journal} {Phys. Rev. Lett.}\ }\textbf {\bibinfo {volume} {106}},\ \bibinfo
  {pages} {040403} (\bibinfo {year} {2011}{\natexlab{a}})}\BibitemShut
  {NoStop}%
\bibitem [{\citenamefont {Lloyd}\ \emph
  {et~al.}(2011{\natexlab{b}})\citenamefont {Lloyd}, \citenamefont {Maccone},
  \citenamefont {Garcia-Patron}, \citenamefont {Giovannetti},\ and\
  \citenamefont {Shikano}}]{Lloyd2011PRD}%
  \BibitemOpen
  \bibfield  {author} {\bibinfo {author} {\bibfnamefont {S.}~\bibnamefont
  {Lloyd}}, \bibinfo {author} {\bibfnamefont {L.}~\bibnamefont {Maccone}},
  \bibinfo {author} {\bibfnamefont {R.}~\bibnamefont {Garcia-Patron}}, \bibinfo
  {author} {\bibfnamefont {V.}~\bibnamefont {Giovannetti}}, \ and\ \bibinfo
  {author} {\bibfnamefont {Y.}~\bibnamefont {Shikano}},\ }\href {\doibase
  10.1103/PhysRevD.84.025007} {\bibfield  {journal} {\bibinfo  {journal} {Phys.
  Rev. D}\ }\textbf {\bibinfo {volume} {84}},\ \bibinfo {pages} {025007}
  (\bibinfo {year} {2011}{\natexlab{b}})}\BibitemShut {NoStop}%
\bibitem [{\citenamefont {Svetlichny}(2011)}]{Svetlichny2011}%
  \BibitemOpen
  \bibfield  {author} {\bibinfo {author} {\bibfnamefont {G.}~\bibnamefont
  {Svetlichny}},\ }\href {\doibase 10.1007/s10773-011-0973-x} {\bibfield
  {journal} {\bibinfo  {journal} {Int. J. Theor. Phys.}\ }\textbf {\bibinfo
  {volume} {50}},\ \bibinfo {pages} {3903} (\bibinfo {year}
  {2011})}\BibitemShut {NoStop}%
\bibitem [{\citenamefont {Brun}\ and\ \citenamefont
  {Wilde}(2012)}]{BrunWilde2011}%
  \BibitemOpen
  \bibfield  {author} {\bibinfo {author} {\bibfnamefont {T.~A.}\ \bibnamefont
  {Brun}}\ and\ \bibinfo {author} {\bibfnamefont {M.~M.}\ \bibnamefont
  {Wilde}},\ }\href {\doibase 10.1007/s10701-011-9601-0} {\bibfield  {journal}
  {\bibinfo  {journal} {Found. Phys.}\ }\textbf {\bibinfo {volume} {42}},\
  \bibinfo {pages} {341} (\bibinfo {year} {2012})}\BibitemShut {NoStop}%
\bibitem [{\citenamefont {Ralph}(2007)}]{Ralph2007}%
  \BibitemOpen
  \bibfield  {author} {\bibinfo {author} {\bibfnamefont {T.~C.}\ \bibnamefont
  {Ralph}},\ }\href {\doibase 10.1103/PhysRevA.76.012336} {\bibfield  {journal}
  {\bibinfo  {journal} {Phys. Rev. A}\ }\textbf {\bibinfo {volume} {76}},\
  \bibinfo {pages} {012336} (\bibinfo {year} {2007})}\BibitemShut {NoStop}%
\bibitem [{\citenamefont {Ralph}\ \emph {et~al.}(2009)\citenamefont {Ralph},
  \citenamefont {Milburn},\ and\ \citenamefont {Downes}}]{Ralph2009}%
  \BibitemOpen
  \bibfield  {author} {\bibinfo {author} {\bibfnamefont {T.~C.}\ \bibnamefont
  {Ralph}}, \bibinfo {author} {\bibfnamefont {G.~J.}\ \bibnamefont {Milburn}},
  \ and\ \bibinfo {author} {\bibfnamefont {T.}~\bibnamefont {Downes}},\ }\href
  {\doibase 10.1103/PhysRevA.79.022121} {\bibfield  {journal} {\bibinfo
  {journal} {Phys. Rev. A}\ }\textbf {\bibinfo {volume} {79}},\ \bibinfo
  {pages} {022121} (\bibinfo {year} {2009})}\BibitemShut {NoStop}%
\bibitem [{\citenamefont {Wallman}\ and\ \citenamefont
  {Bartlett}(2012)}]{Wallman2012}%
  \BibitemOpen
  \bibfield  {author} {\bibinfo {author} {\bibfnamefont {J.~J.}\ \bibnamefont
  {Wallman}}\ and\ \bibinfo {author} {\bibfnamefont {S.~D.}\ \bibnamefont
  {Bartlett}},\ }\href {\doibase 10.1007/s10701-012-9635-y} {\bibfield
  {journal} {\bibinfo  {journal} {Found. Phys.}\ }\textbf {\bibinfo {volume}
  {42}},\ \bibinfo {pages} {656} (\bibinfo {year} {2012})}\BibitemShut
  {NoStop}%
\bibitem [{\citenamefont {Cavalcanti}\ and\ \citenamefont
  {Menicucci}(2010)}]{Cavalcanti2010}%
  \BibitemOpen
  \bibfield  {author} {\bibinfo {author} {\bibfnamefont {E.~G.}\ \bibnamefont
  {Cavalcanti}}\ and\ \bibinfo {author} {\bibfnamefont {N.~C.}\ \bibnamefont
  {Menicucci}},\ }\href@noop {} {\  (\bibinfo {year} {2010})},\ \Eprint
  {http://arxiv.org/abs/1004.1219} {arXiv:1004.1219 [quant-ph]} \BibitemShut
  {NoStop}%
\bibitem [{\citenamefont {Ralph}\ and\ \citenamefont
  {Myers}(2010)}]{RalphMyers2010}%
  \BibitemOpen
  \bibfield  {author} {\bibinfo {author} {\bibfnamefont {T.~C.}\ \bibnamefont
  {Ralph}}\ and\ \bibinfo {author} {\bibfnamefont {C.~R.}\ \bibnamefont
  {Myers}},\ }\href {\doibase 10.1103/PhysRevA.82.062330} {\bibfield  {journal}
  {\bibinfo  {journal} {Phys. Rev. A}\ }\textbf {\bibinfo {volume} {82}},\
  \bibinfo {pages} {062330} (\bibinfo {year} {2010})}\BibitemShut {NoStop}%
\bibitem [{\citenamefont {Allen}(2014)}]{Allen2014}%
  \BibitemOpen
  \bibfield  {author} {\bibinfo {author} {\bibfnamefont {J.-M.~A.}\
  \bibnamefont {Allen}},\ }\href {\doibase 10.1103/PhysRevA.90.042107}
  {\bibfield  {journal} {\bibinfo  {journal} {Phys. Rev. A}\ }\textbf {\bibinfo
  {volume} {90}},\ \bibinfo {pages} {042107} (\bibinfo {year}
  {2014})}\BibitemShut {NoStop}%
\bibitem [{\citenamefont {{Dong}}\ \emph {et~al.}(2017)\citenamefont {{Dong}},
  \citenamefont {{Chen}},\ and\ \citenamefont {{Zhou}}}]{DongChenZhou2017}%
  \BibitemOpen
  \bibfield  {author} {\bibinfo {author} {\bibfnamefont {X.}~\bibnamefont
  {{Dong}}}, \bibinfo {author} {\bibfnamefont {H.}~\bibnamefont {{Chen}}}, \
  and\ \bibinfo {author} {\bibfnamefont {L.}~\bibnamefont {{Zhou}}},\
  }\href@noop {} {\  (\bibinfo {year} {2017})},\ \Eprint
  {http://arxiv.org/abs/1711.06814} {arXiv:1711.06814 [quant-ph]} \BibitemShut
  {NoStop}%
\bibitem [{Note1()}]{Note1}%
  \BibitemOpen
  \bibinfo {note} {We thank an anonymous referee for providing us this proof.}\BibitemShut {Stop}%
\bibitem [{\citenamefont {{Nath Bera}}\ \emph {et~al.}(2017)\citenamefont
  {{Nath Bera}}, \citenamefont {{Riera}}, \citenamefont {{Lewenstein}},\ and\
  \citenamefont {{Winter}}}]{Bera2017}%
  \BibitemOpen
  \bibfield  {author} {\bibinfo {author} {\bibfnamefont {M.}~\bibnamefont
  {{Nath Bera}}}, \bibinfo {author} {\bibfnamefont {A.}~\bibnamefont
  {{Riera}}}, \bibinfo {author} {\bibfnamefont {M.}~\bibnamefont
  {{Lewenstein}}}, \ and\ \bibinfo {author} {\bibfnamefont {A.}~\bibnamefont
  {{Winter}}},\ }\href@noop {} {\  (\bibinfo {year} {2017})},\ \Eprint
  {http://arxiv.org/abs/1707.01750} {arXiv:1707.01750 [quant-ph]} \BibitemShut
  {NoStop}%
\bibitem [{\citenamefont {Horodecki}\ \emph {et~al.}(2005)\citenamefont
  {Horodecki}, \citenamefont {Horodecki}, \citenamefont {Sen(De)},\ and\
  \citenamefont {Sen}}]{Horodecki2005}%
  \BibitemOpen
  \bibfield  {author} {\bibinfo {author} {\bibfnamefont {M.}~\bibnamefont
  {Horodecki}}, \bibinfo {author} {\bibfnamefont {R.}~\bibnamefont
  {Horodecki}}, \bibinfo {author} {\bibfnamefont {A.}~\bibnamefont {Sen(De)}},
  \ and\ \bibinfo {author} {\bibfnamefont {U.}~\bibnamefont {Sen}},\ }\href
  {\doibase 10.1007/s10701-005-8661-4} {\bibfield  {journal} {\bibinfo
  {journal} {Found. Phys.}\ }\textbf {\bibinfo {volume} {35}},\ \bibinfo
  {pages} {2041} (\bibinfo {year} {2005})}\BibitemShut {NoStop}%
\bibitem [{\citenamefont {Hulpke}\ \emph {et~al.}(2006)\citenamefont {Hulpke},
  \citenamefont {Poulsen}, \citenamefont {Sanpera}, \citenamefont {Sen(De)},
  \citenamefont {Sen},\ and\ \citenamefont {Lewenstein}}]{Hulpke2006}%
  \BibitemOpen
  \bibfield  {author} {\bibinfo {author} {\bibfnamefont {F.}~\bibnamefont
  {Hulpke}}, \bibinfo {author} {\bibfnamefont {U.~V.}\ \bibnamefont {Poulsen}},
  \bibinfo {author} {\bibfnamefont {A.}~\bibnamefont {Sanpera}}, \bibinfo
  {author} {\bibfnamefont {A.}~\bibnamefont {Sen(De)}}, \bibinfo {author}
  {\bibfnamefont {U.}~\bibnamefont {Sen}}, \ and\ \bibinfo {author}
  {\bibfnamefont {M.}~\bibnamefont {Lewenstein}},\ }\href {\doibase
  10.1007/s10701-005-9035-7} {\bibfield  {journal} {\bibinfo  {journal} {Found.
  Phys.}\ }\textbf {\bibinfo {volume} {36}},\ \bibinfo {pages} {477} (\bibinfo
  {year} {2006})}\BibitemShut {NoStop}%
  \bibitem [{\citenamefont {{Monsel}}\ \emph {et~al.}(2018)\citenamefont
  {{Monsel}}, \citenamefont {{Elouard}},\ and\ \citenamefont
  {{Auff{\`e}ves}}}]{Monsel2018}%
  \BibitemOpen
  \bibfield  {author} {\bibinfo {author} {\bibfnamefont {J.}~\bibnamefont
  {{Monsel}}}, \bibinfo {author} {\bibfnamefont {C.}~\bibnamefont {{Elouard}}},
  \ and\ \bibinfo {author} {\bibfnamefont {A.}~\bibnamefont {{Auff{\`e}ves}}},\
  }\href@noop {} {\  (\bibinfo {year} {2018})},\ \Eprint
  {http://arxiv.org/abs/1804.02296} {arXiv:1804.02296 [quant-ph]} \BibitemShut
  {NoStop}%
\bibitem [{\citenamefont {Batalh\~ao}\ \emph {et~al.}(2015)\citenamefont
  {Batalh\~ao}, \citenamefont {Souza}, \citenamefont {Sarthour}, \citenamefont
  {Oliveira}, \citenamefont {Paternostro}, \citenamefont {Lutz},\ and\
  \citenamefont {Serra}}]{Batalh2015}%
  \BibitemOpen
  \bibfield  {author} {\bibinfo {author} {\bibfnamefont {T.~B.}\ \bibnamefont
  {Batalh\~ao}}, \bibinfo {author} {\bibfnamefont {A.~M.}\ \bibnamefont
  {Souza}}, \bibinfo {author} {\bibfnamefont {R.~S.}\ \bibnamefont {Sarthour}},
  \bibinfo {author} {\bibfnamefont {I.~S.}\ \bibnamefont {Oliveira}}, \bibinfo
  {author} {\bibfnamefont {M.}~\bibnamefont {Paternostro}}, \bibinfo {author}
  {\bibfnamefont {E.}~\bibnamefont {Lutz}}, \ and\ \bibinfo {author}
  {\bibfnamefont {R.~M.}\ \bibnamefont {Serra}},\ }\href {\doibase
  10.1103/PhysRevLett.115.190601} {\bibfield  {journal} {\bibinfo  {journal}
  {Phys. Rev. Lett.}\ }\textbf {\bibinfo {volume} {115}},\ \bibinfo {pages}
  {190601} (\bibinfo {year} {2015})}\BibitemShut {NoStop}%
\bibitem [{\citenamefont {Brian}(2004)}]{Greene2004}%
  \BibitemOpen
  \bibfield  {author} {\bibinfo {author} {\bibfnamefont {G.}~\bibnamefont
  {Brian}},\ }\href@noop {} {\emph {\bibinfo {title} {Freeman Dyson in : The
  Fabric of the Cosmos: Space, Time, and the Texture of Reality}}}\ (\bibinfo
  {publisher} {Alfred Knopf},\ \bibinfo {year} {2004})\BibitemShut {NoStop}%
\bibitem [{\citenamefont {Krauss}\ and\ \citenamefont
  {Wilczek}(2014)}]{KraussWilczek2014}%
  \BibitemOpen
  \bibfield  {author} {\bibinfo {author} {\bibfnamefont {L.~M.}\ \bibnamefont
  {Krauss}}\ and\ \bibinfo {author} {\bibfnamefont {F.}~\bibnamefont
  {Wilczek}},\ }\href {\doibase 10.1103/PhysRevD.89.047501} {\bibfield
  {journal} {\bibinfo  {journal} {Phys. Rev. D}\ }\textbf {\bibinfo {volume}
  {89}},\ \bibinfo {pages} {047501} (\bibinfo {year} {2014})}\BibitemShut
  {NoStop}%

\end{thebibliography}
\end{document}